\documentclass[groupedaddress]{revtex4}
\usepackage{graphicx}  
\usepackage{dcolumn}   
\usepackage{bm}        
\usepackage[english]{babel}
\usepackage{amsfonts,amsmath,amssymb,mathrsfs}
\usepackage{epstopdf}
\usepackage{subfigure}
\usepackage{color}

\def\a{\alpha}
\def\r{\rho}
\def\s{\sigma}
\def\t{\tau}
\def\m{\mu}
\def\n{\nu}
\def\k{\kappa}
\def\th{\theta}
\def\g{\gamma}\def\G{\Gamma}
\def\L{\Lambda}\def\l{\lambda}
\def\D{\Delta}
\def\la{\langle}
\def\ra{\rangle}
\def\o{\omega}\def\O{\Omega}
\def\d{\delta}
\def\p{\partial}

\def\half{\textstyle{\frac{1}{2}}}

\def\bdoc{\begin{document}}
\def\edoc{\end{document}}

\def\beq{\begin{equation}}
\def\eeq{\end{equation}}
\def\bea{\begin{eqnarray}}
\def\eea{\end{eqnarray}}
\def\ben{\begin{enumerate}}
\def\een{\end{enumerate}}
\def\la{\langle}
\def\ra{\rangle}
\def\a{\alpha}
\def\b{\beta}
\def\g{\gamma}\def\G{\Gamma}
\def\d{\delta}\def\D{\Delta}
\def\e{\epsilon}

\def\th{\theta}
\def\k{\kappa}
\def\l{\lambda}
\def\m{\mu}
\def\n{\nu}
\def\o{\omega}
\def\p{\pi}
\def\r{\rho}
\def\s{\sigma}
\def\t{\tau}
\def\L{{\cal L}}
\def\S{\Sigma }
\def\gsim{\; \raisebox{-.8ex}{$\stackrel{\textstyle >}{\sim}$}\;}
\def\lsim{\; \raisebox{-.8ex}{$\stackrel{\textstyle <}{\sim}$}\;}
\def\gtrsim{\gsim}
\def\lessim{\lsim}
\def\loc{{\rm local}}
\def\vm{v_{\rm max}}
\def\bh{\bar{h}}
\def\del{\partial}
\def\nab{\nabla}
\def\half{{\textstyle{\frac{1}{2}}}}
\def\fourth{{\textstyle{\frac{1}{4}}}}

\def\bD{{\bf D}}
\def\bE{{\bf E}}
\def\bF{{\bf F}}
\def\bB{{\bf B}}
\def\bP{{\bf P}}
\def\bV{{\bf v}}
\def\bv{{\bf v}}
\def\bx{{\bf x}}
\def\by{{\bf y}}
\def\bz{{\bf z}}
\def\ba{{\bf a}}
\def\bd{{\bf d}}
\def\bs{{\bf s}}
\def\bn{{\bf n}}
\def\bp{{\bf p}}

\def\O{\Omega}

\def\br{{\bf r}}
\def\bnab{{\bf \nab}}

\def\tE{\tilde{E}}
\def\tL{\tilde{L}}

\begin{document}

\title{Might black holes reveal their inner secrets?}
\author{Ted Jacobson$^*$ and Thomas P. Sotiriou$^{\dagger}$}
\affiliation{$^*$Center for Fundamental Physics,  University of Maryland, College Park, MD 20742-4111, USA}
\affiliation{$^\dagger$Department of Applied Mathematics and Theoretical Physics, Centre for  
Mathematical Sciences, University of Cambridge, Wilberforce Road,  
Cambridge, CB3 0WA, UK}
\date{\today} 
\begin{abstract}
{\em Black holes harbor a spacetime singularity of infinite curvature, 
where classical spacetime physics breaks down, and 
current theory cannot predict what will happen. However, 
the singularity is invisible from the outside because
strong gravity traps all signals, even light, behind an event horizon. 
In this essay we discuss  
whether it might 
be possible to destroy the horizon,  if  
a body is tossed into the black hole
so as to make it spin faster  and/or have more charge than a certain limit. 
It turns out that one could expose a ``naked" singularity
if effects of the body's own gravity can be neglected.
We suspect however that such neglect is unjustified.}
\end{abstract}  
\maketitle


\subsection*{ \em Black holes and cosmic censorship}
Black holes are among the most fascinating concepts in physics. 
The notion of a gravitating body from which light cannot escape
can be traced all the way back to  geologist John C. Michell in 1783, 
but the modern concept to which the term black hole refers 
arises only within Einstein's 1915 general relativity theory of gravity. 
The understanding of these objects came much later, 
and the term ``black hole" came into use only in the 1960s, 
popularized by John Wheeler.\\

What exactly is a black hole? Michell's Newtonian conception of an 
object whose escape velocity exceeds the speed of light provides a 
rough understanding of the black hole size, but is qualitatively 
deeply inaccurate. It describes a situation in which light would rise
to a height depending on its initial launch location, and then fall back
toward the object. The black hole of general relativity is quite different
since, as in special relativity, the speed of light 
is independent of the conditions of the source of the light. 
The critical concept needed to describe a black hole is that of
the event horizon: a one-way boundary into which objects and light 
can fall, but out of which nothing, including light or any signal, 
can come --- a boundary in spacetime beyond which events cannot affect an outside observer. 
Light does not rise up to the horizon and fall back, rather, light
at the horizon simply hovers at the horizon.
It is the presence of the horizon that motivates the name ``black hole".
The horizon itself is just a ``surface of no return"
in the spacetime, and locally nothing special is happening there.
What lies inside the horizon
is a wholly different matter however.
The black hole 
harbors a spacetime singularity of infinite curvature --- a location 
where the tidal accelerations characterizing the gravitational field become infinite.\\

The presence of divergences at the singularity signals the breakdown of general relativity. Einstein's theory would be unable to predict the outcome of events in the vicinity of the singularity, and unusual phenomena could take place there. To describe these would presumably require a theory that merges gravity with quantum phenomena, a theory has not yet been fully formulated much less understood. \\

But are spacetime singularities physically relevant or are they simply mathematical peculiarities of special solutions to Einstein's theory? As a matter of fact, Penrose and Hawking have managed to show that singularities are not only relevant but actually inevitable in gravitational collapse \cite{Penrose:1964wq,Hawking:1969sw}. Chandrasekhar at first \cite{chandra} and later others had already shown that for spherical matter configurations, gravitational collapse most likely leads to singularities provided that the initial mass is large enough. The importance of Penrose and Hawking's results lies in their generality, as 
they do not rely on specific solutions of Einstein's equations or special symmetry requirements. 
\\

So, the collapse of some dying stars, for instance, will inevitably lead to the formation of a black hole containing a spacetime singularity. In this case the presence of the event horizon conveniently hides the singularity from our curious eyes and relieves us from the burden of having to worry about it
if all we wish to understand is observations from outside. 
It would actually be much more exciting if there were no horizon: a visible singularity would give us observational access to the unknown phenomenology mentioned previously. Hence the pertinent question, 
does an event horizon always hide any singularities?
As Roger Penrose first put it in 1969~\cite{Penrose:1969pc}, 
``does there exist a Òcosmic
censorÓ who forbids the appearance of naked singularities, 
clothing each one in an absolute event horizon?" The conjecture that
such a censor indeed exists is called the ``cosmic censorship conjecture".
\\

The physics of event horizons, unlike that of singularities,  
should be well described by classical general relativity, 
so cosmic censorship can be tested and possibly proven without going
beyond known physics.  Evidence 
from several directions suggests that any singularity arising to the 
future of generic (not infinitely finely tuned)
non-singular initial conditions, may indeed always be hidden behind a black hole 
event horizon~\cite{Wald:1997wa,Penrose:1999vj}. However, we are far from
having a definitive proof. 
\\

Much of the evidence in favor of cosmic censorship comes from
the failure of attempts to violate it in thought experiments. Given the difficulty
of a direct attack on the general question, this strategy
continues to offer an attractive approach to the problem. 
Even in the context of simplifying approximations, 
such thought experiments can uncover mechanisms
tending to uphold censorship, and they can produce
scenarios where censorship would be violated if the approximation
were valid. In the latter case, they focus our attention on
those ``dangerous" scenarios and on the limits of validity 
of the approximation scheme. This type of approach is what
we will be discussing here.\\

Before going further, it is interesting to mention the implications 
of naked singularities for black hole thermodynamics, 
the extension of ordinary thermodynamics to systems containing black holes \cite{waldreview}. 
According to black hole thermodynamics the area of the horizon of the black hole 
is proportional to its entropy. 
In analogy with the second law of ordinary thermodynamics, the area of a black hole can never 
decrease in a classical process. The role of temperature is played by what is called surface gravity. 
In analogy with the third law of ordinary thermodynamics the surface gravity cannot be reduced 
to zero in a finite time. Hawking has provided a proof of
the second law of black hole mechanics~\cite{Hawking:1971tu}, and Israel a 
proof of the third law~\cite{Israel}. These proofs assume, among other things, that there
are no naked singularities.  A violation of cosmic censorship would invalidate both proofs.
Moreover, if horizons could be ``destroyed'' by processes like those discussed in the following,
the foundations of black hole thermodynamics would seem to crumble.
So there may well be some deep connection between the validity of cosmic censorship 
and the thermodynamics of spacetime~\cite{tos}.

\subsection*{\em Tilting at windmills}

A much studied 
obvious strategy for ``creating'' a naked singularity is to attempt to ``destroy" a black hole horizon,
that is, to strip the singularity of its clothing. In order to understand this better let us return to the basics of black holes. In general relativity black holes are fully characterized by three quantities: their mass $M$, spin angular momentum $J$, and electric charge $Q$. The spacetime in the vicinity of the black hole is described by the Kerr--Newman (K-N) metric --- the measure of distances and times --- which contains the three parameters $M$, $J$ and $Q$. The K-N metric describes a black hole as long as the mass is sufficiently large compared to a combination of the charge and angular momentum, 
$M^2 \ge a^2+Q^2$, where $a=J/M$. (We adopt units with Newton's constant $G$ 
and the speed of light $c$ both set equal to unity.
Displaying hidden factors of $G$ and $c$, the quantities $MG/c^2$, $a/c$ and $Q\sqrt{G}/c^2$
all have the dimension of length.) 
The case where $M^2 = a^2 + Q^2$ is called an extremal black hole,
while for $M^2 < a^2+Q^2$ there is no event horizon and the K-N metric actually describes a naked singularity. Therefore, it would naively seem that to create a naked singularity all one need do is to start with a black hole and toss in matter with enough angular momentum or charge so as to drive its parameters beyond the extremal limit, leaving it no option but to expose the singularity.\\

On further thought there are some subtleties involved in such a scenario. 
First of all, the K-N metric describes stationary configurations,
so the proposed strategy can work as stated only if, after having absorbed the matter, 
the system settles down to a stationary configuration containing all the mass, angular
momentum and charge, i.e. without having shed the excess angular momentum
or charge in the settling down process. Such an outcome is by no means
guaranteed. Indeed, it seems rather unlikely, given the evidence that
the trans-extremal K-N metric is unstable~\cite{Dotti:2006gc,Dotti:2008js}. 
If such instability occurs (the uncertainty lies 
in the proper boundary conditions at the singularity), the system may well not settle down
to such a metric, and at present nobody knows what it would do. 
What this means is that to demonstrate the creation of a naked singularity one
would have to follow the evolution further than the initial ``absorption" of the 
extra matter. So the possibility of initially 
overspinning or overcharging an initial black hole configuration
can only be taken as an indication that cosmic censorship {\it might} fail.\\
 
A second subtlety is that the notion of ``exposing" the singularity may be 
inappropriate, since the singularity inside a perturbed charged or rotating stationary
black hole cannot send signals to any point, even those interior to the horizon.
(Here we assume that the Cauchy horizon inside the black 
hole is indeed unstable, as evidence indicates~\cite{Cauchy}.)
That is, it has no nonsingular future. 
Hence it is not so clear that, if a horizon could be 
``destroyed", the result would be to expose the singularity that would have been 
there had the horizon {\it not} been destroyed.  It might produce
a wholly different singularity.  Nevertheless, either way 
the process would violate cosmic censorship.\\

To follow the evolution exactly is a very difficult problem that presumably requires 
numerical solution of the Einstein equation. The difficulty arises because 
general relativity is a highly non-linear theory, but even linearizing around a background
black hole solution remains a very difficult problem. So most studies of cosmic
censorship to date have been carried out in the simple framework of the ``test-body 
approximation" in which the matter being tossed into the black hole 
moves in a way determined by the original black hole metric (and electric field, if present).
This approximation does not   
take into account effects arising from the gravitation of the
matter itself, namely the gravitational radiation and the self-force.
To try to justify the test-body approximation one can assume the conditions
\beq
\label{small}
\delta E\ll M, \quad  \delta J\ll M^2, \quad \delta Q\ll M,
\eeq
where $\delta E$, $\delta J$, and $\delta Q$  denote the energy, angular momentum and charge 
of the body. \\

Now, provided the body can be tossed into the black hole, 
the final composite object would have mass $M+\delta E$, 
angular momentum $J+\delta J$ and charge  $Q+\delta Q$. 
In order for the K-N metric with these parameters to be a naked singularity they would have to satisfy the inequality
\beq
\label{jqover}
(M+\delta E)^2 < \left(\frac{J+\delta J}{M+\delta E}\right)^2+(Q+\delta Q)^2.
\eeq
\\
Various special cases have been considered in the literature. For instance,
one could consider only charge or only angular momentum.
Alternatively, one could start with charge but no angular momentum 
for the black hole, and then try to reach a trans-extremal state by adding 
angular momentum but no charge, or vice versa.\\

In nature, black holes tend to increase their angular momentum by accreting matter, 
whereas they tend to decrease their charge by attracting opposite and repelling same charges. For a point of principle, one may perhaps ignore these facts, but 
nevertheless it would be much more provocative and promising if arguments
showed that a naked singularity could be created using only angular momentum,
since that might then actually occur in nature. 
Therefore, we choose to focus here mostly 
on the spinning black hole case.
We shall demonstrate that in the test body approximation a trans-extremal 
condition can be attained, even when taking into account constraints on the
size and structure of the body. The limitations of the test-body aproximation
will then be addressed.\\

With $Q=\delta Q=0$, the inequality in eq.~(\ref{jqover})  takes the form
\beq
J+\d J > (M+\d E)^2.
\eeq
This yields a lower bound on the required angular momentum carried by the body, for a given energy $\delta E$:
\beq\label{over-spin}
\d J > \d J_{min}=(M^2-J)+2M \d E+\d E^2.
\eeq
Since we are assuming $\d E\ll M$, it might seem
that the $\d E^2$ term may as well just be neglected
at this stage. However, as we will see later on, the presence of that term imposes 
an upper bound on $\d E$ and $\d J$ and, therefore, should not be neglected. 
\\

We can already extract a useful piece of information from this last equation. Dividing both side by $\delta E^2$ and observing that each term on the right hand side should by itself be smaller that the left hand side, we get
\beq\label{largespin}
\d J/\d E^2>2M/\d E \gg1.
\eeq
If $\d E$ were equal to the rest mass of the
body (it can be much less if the body is deeply bound by the gravitational field of the black hole), 
and if $\d J$ comes from spin (rather than orbital angular momentum),
this would imply that the body
has angular momentum far over the 
extremal ratio. 
In that case
the body could not be a black hole. This does not mean 
that it would have to be a naked singularity itself, 
as there is no a priori upper limit to this ratio for bodies other 
than black holes. Stars for instance can easily have ratios much larger than 1.\\

We have got a lower bound on the angular momentum of the body from 
the requirement that the composite object be a naked singularity. 

An upper bound comes from the requirement
that the body does indeed
cross the horizon and 
end up in the black hole --- or rather what was formerly the black hole. 
To check whether
this requirement is satisfied one may resort to the equations of motion for the 
body~\cite{Wald}. These are the Papapetrou equation if the body's angular momentum is due to spin, 
and the geodesic equation if it is orbital angular momentum. 
But a simpler and more transparent 
method is to just consider the flux of energy and angular momentum into the black hole when the body falls across the horizon. The requirement that
the energy momentum tensor satisfies the null energy condition (which follows for
example if the energy density is positive in all local reference frames)
yields \cite{tedthomas}
 \beq\label{go-in2}
\d E\ge\O_H\d J,
\eeq
where $\O_H=a/2Mr_+$ is the angular velocity of the horizon 
and $r_+=M+(M^2-a^2)^{1/2}$ is the horizon radius
in Boyer-Lindquist 
coordinates. This condition can be written as
 \beq\label{go-in}
\d J \le \d J_{max}=\frac{2Mr_+}{a} \d E.
\eeq
It guarantees that the body can 
fall across the horizon starting from {\it some} point outside,
although in general 
the body is in a bound orbit that does not come 
from spatial infinity.\\

We now have both an upper and a lower bound for the 
angular momentum of the body, for a given energy. 
As long as $\d J_{min}<\d J_{max}$ for some
$\d E$, there will 
be values of $\d J$ and $\d E$ satisfying both inequalities
(\ref{over-spin}) and (\ref{go-in}). First let us 
suppose the black hole starts exactly extremal, i.e. 
$J=M^2$. Then $a=M=r_+$, so one has
$\d J_{min}=2M\d E +\d E^2$ 
and $\d J_{max}=2M\d E$. 
This implies that $\delta J_{min}> \delta J_{max}$ 
so it is impossible to over-spin
the black hole. This was observed long ago by 
Wald~\cite{Wald} (via a significantly more involved calculation). 
The physical interpretation is the following: In the case of the spinning 
body, the spin-spin interaction with the spin of the black hole is sufficiently repulsive
to prevent the body from falling in if it would have overspun the black hole. 
In the orbital angular momentum case, the impact parameter of the body is too large for it to 
hit the horizon if it has the angular momentum required to overspin.\\

In the sub-extremal case, however,
the inequalities {\it can} be satisfied, as was shown in \cite{tedthomas}. 
The limiting case where the body is dropped from a point on the horizon 
had been considered previously by Hod \cite{Hod:2002pm}. 
To understand the range of overspinning parameters, 
it is helpful to visualize the inequalities
graphically. If $\d J_{max}$ and $\d J_{min}$ 
are plotted vs. 
$\d E$, 
the former is a straight line through the
origin, with slope $2M r_+/a>2M$,
while the latter is a parabola with positive
intercept, slope $2M$ at the intercept, and 
curved upwards. Some algebra reveals that 
the parabola always intersects
the straight line in two points. The allowed
values of $\d E$ and $\d J$ are those in the
compact region above the parabola and on or
below the
straight line. Note that
if the $\d E^2$
is neglected in (\ref{over-spin}),
the parabola is replaced by a straight line,
and no upper bound is imposed on the 
allowed values. 
The case considered by Hod~\cite{Hod:2002pm},
i.e.\ that of dropping the particle from a point on the horizon,
corresponds to the upper boundary of this region, $\d J =\d J_{max}$.\\

To quantitatively characterize the overspinning region we can 
expand in the small dimensionless quantity
$\e\ll1$ defined by 
\beq\label{epsilon}
J/M^2=a/M=1-2\e^2.
\eeq
(Hubeny~\cite{Hubeny:1998ga} used the same parameter to analyze the 
charged case, see below.)
The parameter $\e$ measures how close to extremality the 
black hole is to begin with. 
It is now useful to adopt units with $M=1$, 
to keep the expressions simpler. 
Then we have 
\bea\label{minmax}
\d J_{min} &=& 2\e^2+2\d E + \d E^2\\
\d J_{max}&=& (2+4\e)\d E,\label{jmax}
\eea
where terms of order $O(\e^2\d E)$ have been dropped
in (\ref{jmax}).
The allowed range of $\d E$ lies where the difference
\beq\label{max-min}
\d J_{max}-\d J_{min} = -2\e^2+4\e\d E - \d E^2 
\eeq
is positive, i.e.
\beq\label{dEe}
(2-\sqrt{2})\e <\d E<(2+\sqrt{2})\e.
\eeq
In particular, $\d E$ must be of order $\e$,
which is consistent with the requirement
(\ref{small}) that the body make only a small perturbation. 
For a given $\d E$, the allowed values of 
$\d J$ are near $2\d E$, so we must have 
\beq\label{dJ/dE}
\d J\sim \d E.
\eeq
Note that the {\it width} (\ref{max-min}) of the allowed
range of $\d J$ is only of order $\e^2\ll\e$.\\

Clearly, the black hole must start out
very nearly extremal. To be
somewhat more quantitative, according to (\ref{small})
and (\ref{dEe}) we must have $\e\ll1$,
and $a - 1=2\e^2$ is parametrically smaller.
For example, if $\e=10^{-2}$, then 
the initial black hole must have 
$a=0.9998$. For a thought experiment,
we can imagine even smaller values of $\e$.\
We conclude that, if the body can be treated as a
point test particle moving on the background spacetime 
of the black hole, the black hole
can indeed be over-spun!\\

A similar conclusion had been previously reached by 
Hubeny~\cite{Hubeny:1998ga},  
for the case of adding charge to a charged black hole ($J=\delta J = 0$). 
In analogy to the spinning case, in the charged case the two constraints are
\begin{align}
&\delta Q>M- Q+\delta E,\\
&\delta Q \le \frac{r_+}{Q}\delta E,
\end{align}
where now $r_+ = M + \sqrt{M^2-Q^2}$.
If the black hole starts out extremal, $M=Q=r_+$, then overcharging is 
impossible. However if $M>Q$, then $r_+>Q$, and one easily 
sees by visualizing the inequalities graphically (now they are both straight lines)
that there is an infinite range of solutions, once $\d E$ is 
greater than a certain minimum value.\\

A similar conclusion was also reached by de Felice and Yu~\cite{deFelice:2001wj},
for the case of adding angular momentum to an extremally charged
black hole ($Q=M$, $J=\d Q=0$). In this case the minimum $\d J$ to overspin
is given by 
\beq\label{deF}
\d J > (M+\d E)\sqrt{2M\d E + \d E^2},
\eeq
and there is no maximum $\d J$, since the only requirement for the
body to fall across the horizon is $\d E\ge 0$, which does not involve $\d J$. 
Note that to lowest order in $\d E/m$ the minimum overspinning $\d J$
is $\d J_{\rm min} = M\sqrt{2M\d E}$.\\

\subsection*{ \em Does size matter?}

So far we have characterized the body only by its energy 
$\delta E$, angular momentum
$\delta J$ and charge $\d Q$.
We did not consider restrictions placed on its size and structure:
It should
be sufficiently small to justify use of a 
test particle approximation, and 
it should be composed of matter 
having positive energy density and no 
unphysically large stresses. 
The next step is to take into account these issues, which we shall
do here for the case with no charge~\cite{tedthomas}. \\ 

We begin with the spinning case. 
For simplicity we assume that the body 
is dropped along the rotational axis of the black hole.
We first consider the case where the body has $\d E\sim m$,
and is not spinning relativistically, so its spin
angular momentum
is given by $\d J\sim mvR\sim \d E\, v R$, where 
$v$ is the surface velocity and 
$R$ is the equatorial radius. The condition $v<1$ then implies
$R>\d J/\d E$. We saw above that
the ratio $\d J/\d E$ must be of order unity (\ref{dJ/dE}), 
that is
of order $M$. In this case the body must be larger than
the black hole, so it simply will not ``fit" in the transverse
direction, and 
in any case treating it as a point particle with spin
would be unjustified,
since that rests on the
smallness of the size of the body compared to the 
ambient radius of curvature. Moreover, one can 
show that the radial tidal 
stress required to hold the body together 
would be larger than the energy density, violating
energy conditions.
It cannot help to allow ultra-relativistic tangential
velocity:  as a simple Newtonian estimate shows \cite{tedthomas},
that would require unphysical stresses holding the body together.
The conclusion is that it is impossible to 
over-spin the black hole if the body's 
energy is 
close to its rest mass, $\d E\sim m$.
\\


Since the angular momentum involves the
rest mass $m$, not the energy $\delta E$,
it might be possible to achieve a large enough 
$\d J$ with a small enough size $R$, 
without requiring unphysical matter,
by dropping the body from a position
where it is deeply bound, $\d E\ll m$. 
This might be achieved by slowly lowering the
body on a tether,  down to the near the
black hole horizon, before dropping it in.
Now we reconsider whether the size restrictions 
can be met in this setting. \\

We begin with the restrictions on the rest mass $m$. 
If $m$ is much greater than $\d E$, then
the test body approximation requires
that we impose not only $\d E\ll 1\, (=M)$ (\ref{small}),
but also $m\ll 1$. There is also a lower bound on $m$,
coming from an upper bound on $R$: 
the angular momentum is  
$\d J\sim mvR$,
hence
(restricting to nonrelativistic spin $v<1$ as required
by the previously mentioned result)
$R > \d J/m\simeq 4\e/m$.
The requirement $R\ll1$ then yields
$m\gg\e$. The mass and size must therefore fall within
the ranges
\beq\label{bounds}
\e\ll m\ll 1, \qquad 4\e/m\lessim R\ll 1.
\eeq
To these conditions we must add the requirement 
$R\gsim m$
that the body is not a black hole, as explained above.\\

The inequality (\ref{go-in2})
guarantees that the body can cross the horizon with the chosen values
of energy and angular momentum, but since the deeply
bound drop point lies at a finite distance from the horizon it 
is necessary to check that (a) the spinning body would actually fall into the
black hole rather than being repelled, and (b) it is possible to choose 
the polar radius of the
body $R_{\rm polar}$ to be smaller than
the proper distance $d$ from the horizon to the drop point
\beq\label{R<d}
R_{\rm polar}<d,
\eeq
so that it can 
fully ``fit" outside the black hole and be localized at the drop point. \\
In order to fall in, 
the maximum value that $d$ 
can have, given the allowed values of $\delta E$ and $\delta J$, 
turns out~\cite{tedthomas} to be
\beq
d_{max}\simeq \e/m.
\eeq
Thus we arrive at the bound
\beq
R_{\rm polar}<\e/m.
\eeq
Together with (\ref{bounds}), this means that the body
must be at least somewhat oblate, $R_{\rm polar} \lessim R/4$,
but not exceedingly so \footnote{In \cite{tedthomas} the
possibility that  $R_{\rm polar} \ne R_{\rm equator}$ was overlooked, 
so it was 
erroneously 
concluded that no value of $R$ could meet all requirements.}.
We conclude that 
the body can be large enough
to possess the requisite angular momentum with a physically
acceptable stress, and still fit outside the black
hole at the drop point~\footnote{de Felice and Yu made a similar
analysis in the case of dropping the spinning body into an extremal charged black hole,
but they computed the coordinate radius corresponding to $d_{\rm max}$, 
rather than the proper distance. In the extremal case, the proper distance to the 
horizon is infinite in the direction orthogonal to the Killing vector, so there is
apparently no requirement that the body be disk-shaped, contrary to
what was stated in \cite{deFelice:2001wj}.\label{dFY}}.
\\

We now turn our attention to the case
of orbital angular momentum in the equatorial 
plane. Here the issue
is that in order to have the required values of 
$\d E$ and $\d J$, the body might have to be in a bound orbit,
which would have a turning point at a maximum radius.
In that case we would need to require that
the body be small enough to fit outside the horizon at this radius.
Since the body can be no smaller than a black hole
with the same rest mass, it is not clear in advance whether
this requirement could be met. However, as has been shown in \cite{tedthomas}
this size constraint is not an issue, since in fact there are 
suitable
orbits that come in from infinity with no turning point. 
This can be shown numerically, but also analytically by the use of 
the effective potential governing the motion of a test particle in a 
Kerr spacetime (Kerr-Newman with no charge) \cite{tedthomas}.\\

Let us briefly 
consider the size and structure requirements
when attempting to overcharge 
or overspin a 
charged 
black hole.
In the case with no angular momentum 
Hubeny~\cite{Hubeny:1998ga} 
showed that the body 
can have the required charge and mass, with 
low internal stresses and size much smaller than
the black hole. Also, she demonstrated that
there
are 
charged
test particle trajectories that fall from infinity into the black hole. 
Therefore, much like the orbital angular momentum case, 
size
constraints are not
an issue. On the contrary, for spinning particles dropped 
radially with radial spin into 
an extremal charged black hole,
de Felice and Yu~\cite{deFelice:2001wj} 
found that the test body must be bound very close to the horizon.
The same is true for a particle carrying orbital angular momentum
but no spin, as we now show.
The radial motion is governed by the 
equation $\dot{r}^2 + (1-1/r)^2(1+\tilde{L}^2/r^2) = \tilde{E}^2$. Here 
$\dot{r}$ is the derivative of the Reissner-Nordstrom radial coordinate
with respect to the particle proper time, and $\tilde{E}=(\d E)/m$ and $\tilde{L}=(\d J)/m$
are the energy and and angular momentum divided by the particle rest mass $m$,
and we have again set $M=1$.
As mentioned after (\ref{deF}), the overspinning requirement is
$\d J/\d E \gsim \sqrt{2/\d E}\gg1$, hence $\tilde{L}/\tilde{E}\gg1$.
For an unbound orbit $\tilde{E}\ge1$, so we infer that $\tilde{L}\gg1$.
There are therefore two turning points where $\dot{r}=0$. 
To fall into the hole the
particle must lie inside the inner turning point,  
which lies at a radius $r_{\rm inner}$ very close to the horizon, 
where $r_{inner}-1\simeq  \tilde{E}/\tilde{L}  \lessim \sqrt{\d E/2}\ll 1$.
However, although the radial coordinate must be very close to that
of the horizon, the proper distance to the horizon, measured in the 
static frame, is infinite for an exactly extremal black hole. 
Hence, in both the spin and orbital cases, there is no need to further
consider size constraints that might have been imposed by the location
of the turning point (see also footnote [20]).\\

\subsection*{\em Including the body's own gravity}

But what about gravitational wave radiation and self-force, 
which our approximation neglects, as mentioned previously? 
In fact, a body orbiting a black will always lose energy 
and angular momentum in gravitational radiation. 
Additionally, it will always experience self-force effects
due to the ``refraction" of its own field on the background of the black hole.
We saw that, from purely kinematic considerations,
the relation between the energy and angular momentum of the 
dropped body must be very finely tuned: they are both of order $\e$ in
magnitude, but the allowed window for angular momentum, given the 
energy, is only of order $\e^2$. Since the over-spinning process we
found involves a delicate balance, it is certainly possible that, 
although small, gravitational radiation and/or 
self-force
effects may always manage to preclude the over-spinning. \\

Given that the inequalities (\ref{over-spin}) and (\ref{go-in}) 
need only
hold on the horizon, one could imagine that the loss of energy 
and angular momentum 
in gravitational 
radiation could be compensated by 
adjusting the initial conditions. 
In the case of an axially symmetric spinning body 
there is no radiation of angular momentum, so it
should be possible to simply compensate for the energy radiated.
To determine whether compensation is actually possible
in the orbital case requires further investigation.   
More worrisome is whether it is possible to overcome the self force effects.
Note that Hubeny found strong indications that for the charged case the 
electromagnetic self-force might  indeed prevent the overcharging, although her 
calculations were not conclusive \cite{Hubeny:1998ga}. 
The gravitational radiation and self-force effects can in principle be calculated numerically, 
at least in the linearized theory, but the calculation would likely be extremely challenging.   
\\

Another distinct effect that 
might 
prevent the creation of a naked singularity is 
the tides raised on the black hole horizon by the falling body. 
These would be irrelevant for the orbital angular momentum case
since the body falls in from spatial infinity. 
In the spinning body case, 
however, in which  the body is lowered
to the horizon and then dropped, 
the tidal bulge of the horizon might perhaps 
make it impossible for the body
to start out in the exterior
while still satisfying the size constraints.\\

Given the 
existing evidence for cosmic censorship, it seems indeed 
likely that neglected 
gravitational effects will come to its rescue. 
Our analysis suggests a dynamical regime in which
it may be interesting to study these neglected effects. 
It turns out that what is ultimately possible in physics 
hangs very much on the details when it comes to 
black holes...at least until the day that we acquire a deeper understanding
of the uncanny tendency of general relativity toward cosmic censorship.

\subsection*{\em Acknowledgements}
This research was supported in part by
the NSF
under grants PHY-0601800 and PHY-0903572, and by STFC. \\

 \edoc